\documentclass[aps,prb]{revtex4}
\usepackage{color}
\begin{document}
\title{On the phase diagram of UGe$_2$}

\author{V. P. Mineev}

\affiliation{Commissariat \`a l'Energie Atomique,
INAC/SPSMS, 38054 Grenoble, France}
\date{\today}

\begin{abstract}
The magneto-elastic mechanism of development of the first order type instability at the phase 
transition to  the ferromagnet state in itinerant ferromagnet-superconductor UGe$_2$ is discussed. The particular property of this material is the precipitous drop of the critical temperature at pressure increase near 14-15 kbar that drastically increases the temperature of the first order instability in respect to the critical temperature. This effect leading to transformation of the second order type transition to the first order one is determined also by the specific heat  increase in 
the temperature interval of the development of critical fluctuations.  After performing the necessary calculations  and estimations using the parameters characterizing the  properties of UGe$_2$  we argue the effectiveness of this mechanism.
\end{abstract}

\maketitle

\section{Introduction}

The pressure-temperature phase diagram of several weak ferromagnets reveals  the common feature. Namely, the  transition from the paramagnetic to the ferromagnetic states at ambient pressure occurs  by means of the second order phase transition. The phase transition temperature decreases with pressure increase such that it reaches zero value at some critical pressure $P_c$.  At some  pressure interval below  $P_c$ the ordered ferromagnetic moment disappears discontinuously. Thus at high pressures and low temperatures  the ferromagnetic and paramagnetic states are divided by the   first order type transition whereas at higher temperatures and  lower pressures this transition is of the second order. Such type of behavior is typical for
MnSi \cite{Pflei97,Thessieu99,Yu04,Uemura07}, itinerant ferromagnet-superconductor UGe$_2$ 
\cite{Terashima01,Terashima02,Pflei02,Settai02,Harada05,Taufour10}, ZrZn$_2$ \cite{Uhlarz04}. The same behavior has been established  in the ferromagnetic compounds Co(Si$_{1-x}$Se$_x$)$_2$ 
\cite{Goto97}
and (Sr$_{1-x}$Ca$_x$)RuO$_3$ \cite{Uemura07} where the role of governing parameter plays 
the concentration of Se and Ca correspondingly.
Also, there was demonstrated clear evidence  for the first order nature of the ferromagnetic transitions in typical ferromagnets like Ni, Fe and Co.\cite{Yang}

The second order phase transition from paramagnetic to itinerant ferromagnetic  state is usually
considered in frame of Stoner theory  which is particular realization of the Landau theory of phase transitions.
The origin of the ferromagnetic transition of the  first order has been attributed to the non-analytic corrections to the free energy  caused by spin-fluctuations calculated at $T=0$ .\cite{Belitz}
The existence of such a type non-analytic corrections for electron-phonon interaction has been first pointed out by G.M.Eliashberg
\cite{Eli63} and then revealed in frame of the Fermi theory in several theoretical papers.  
There was pointed out  \cite{Belitz} that  in a certain parameter regime, the first order transition is 
unstable with respect to a fluctuation induced second-order transition.

 The applicability of this approach to the real itinerant ferromagnetic metals is doubtful. Firstly, there was pointed out the
problem of instability in 
the treatment of the itinerant ferromagnetism in frame isotropic Fermi liquid model 
\cite{Min}.  Secondly, in real material the soft particle-hole excitations interact to the magnetization
fluctuations, presented by not just  Fermi liquid paramagnon excitations but the host crystal magnetization vibrations. The latter  breaks the simple Fermi liquid description of itinerant ferromagnetism in real
materials part of which are strongly anisotropic (Ising-like) ferromagnets with strong spin-orbital coupling.

Another  approach  has been developed in the paper \cite{Yamada} based on the assumption
that the magneto-elastic interaction induces the sign change of the coefficient   in the forth order term  in the Landau free energy expansion in powers of magnetization. This treatment has, however, several ungainly features which we discuss in the Appendix.

Actually the mean field treatment of the magnetoelastic  mechanism   has been put forward earlier 
in the paper \cite{Bean} (see also the phenomenological argumentation \cite{Yang}) where it was demonstrated that the change of transition character from the second to the first order
  takes place at strong enough steepness  of the exchange interaction dependence on interatomic distance and large compressibility.

The magneto-elastic interaction also produces
another general mechanism 
 for instability of second order phase transition 
 toward to the discontinuous  formation of ferromagnetic state from  the paramagnetic one.
 For the first time it was pointed out  by O. K. Rice \cite{Rice} who has demonstrated that at small enough distance from the volume dependent critical temperature $T_c(V)$, where the specific heat $$C_{fl}(\tau) \sim \tau^{-\alpha},$$ $\tau = \frac{T}{Tc(V )} -1$, tends to infinity due to the critical fluctuations ,
the system inverse compressibility $K= -V \frac{\partial P}{\partial V}=V\frac{\partial^2 F}{\partial V^2}$, expressed through the free
energy $F = F_0 + F_{fl}, ~~F_{fl}\sim -T_c\tau^{2-\alpha} $ starts to be negative	
$$K=K_0-A\tau^{-\alpha}\left (\frac{\partial T_c}{\partial V}\right)^2 
=K_0-A_1~C_{fl}(\tau )<	0,$$	
that	contradicts	to thermodynamic stability of the system. 
 It means, that in reality, before there will be reached the temperature corresponding to $K=0$
 the system undergoes  the first order transition, such that to jump over the instability region directly in the ferromagnetic state with finite magnetization and related to it striction deformation. This transition is similar to the jump over the region with $\partial P/\partial V >0$  on the van der Waals isotherm at the liquid-gas transition \cite{StatPhys}. 
 Thus, if  in the system with the fixed volume the phase transition is of the second order with the infinite increase of specific heat then 
  the effect of finite compressibility under assumption that the critical temperature is the only volume dependent  parameter transforms it into the phase transition of the first order.  
In reality,  the striction interaction can change the shape of the free energy singularity in respect to its form in incompressible case. It means the conclusion done in paper \cite{Rice} has no general
character.

The investigation of modifications of the singular part of the free energy near the phase transition   introduced by the striction interaction of the order parameter fluctuations with the acoustic phonons  was undertaken in several papers \cite{Pikin,Shneer}. 
It was shown that the main conclusion about the change of the order of phase transition from the second to the first order still valid. 
Usually, however, it is difficult to reveal the first order character of the phase transition because 
it has quite small latent heat and appears as nearly second order phase transition. The more favorite situation occurs in UGe$_2$. This compound is characterized by the fast enough suppression of the Curie temperature by pressure \cite{Saxena}  that essentially increases the strength of jump.

   In that follows, we estimate the temperature interval where fluctuation corrections to specific heat 
   near the ferromagnetic phase transition in uniaxial ferromagnet are important. Then making use the basic result of the paper \cite{Pikin} we will  demonstrate how it works in concrete case of UGe$_2$ where the critical temperature $T_c(P)$ drops to the zero. 

\section{The fluctuation correction to the specific heat near the Curie temperature}

UGe$_2$ is orthorhombic crystal with ferromagnetic order at ambient pressure found below $T_C=53~K$. 
Magnetic measurements reveal a very strong magnetocrystalline anisotropy \cite{Onuki} with ${\bf a}$ being the easy axis.  We shall denote it as $z$ direction.
The free energy  of strongly anisotropic ferromagnet can be written in terms of one component scalar order parameter corresponding to magnetization density  $M_z({\bf r})$ along $z$ axis. In that follows we shall omit the order parameter index z.
\begin{equation}
{\cal F}=\int d^3{\bf r}\left\{\alpha M^2+ \beta M^4+\gamma_{ij}\nabla_iM\nabla_jM
-\frac{1}{2}\frac{\partial^2M({\bf r})}{\partial z^2}\int\frac{M({\bf r}')d^3{\bf r}'}{|{\bf r}-{\bf r}'|}\right\}
\end{equation}
Here, $$\alpha=\alpha_0(T-T_c)=\alpha_0T_c\tau,\  \  \tau=(T-T_c)/T_c$$ and the gradient terms are written taking into account the orthorhombic anisotropy
$$
\gamma_{ij} = \left(\begin{array}{ccc} \gamma_{xx} & 0 & 0\\
0 & \gamma_{yy} & 0 \\
0 & 0 & \gamma_{yy}
\end{array} \right),
$$
where the $x, y, z$ are directions of the spin axes pinned to $b, c, a$
crystallographic directions correspondingly. The last nonlocal term in Eq. (1) corresponds to magnetostatic energy\cite{Elec,StatPhysII} $-{\bf M}{\bf H}-H^2/8\pi$, where internal magnetic field ${\bf H}$ expressed in terms of magnetization density by means of Maxwell equations $rot{\bf H}=0$ and $div({\bf H}+4\pi{\bf M})=0$.

In that follows we shall use the following estimations for the coefficients in the Landau free energy functional
\begin{eqnarray}
&\alpha_0&=\frac{1}{m^2n},\\
&\beta&=\frac{T_c}{2(m^2n)^2n},\\
\gamma_x\approx&\gamma_y&\approx\gamma_z\approx\frac{T_ca^2}{m^2n}.
\label{3}
\end{eqnarray}
Here, $m=1.4\mu_B$ is the magnetic moment per uranium atom at zero temperature \cite{Ker},
$n=a^{-3}$ is the density of Uranium atoms, which can be approximately taken equal to inverse cube of the nearest-neighbor uranium atoms separation $a=3.85$ Angstrom \cite{Huxley01}.

The  mean field magnetization and the jump of specific heat are
\begin{eqnarray}
&M^2&=-\frac{\alpha}{2\beta}=m^2n\frac{T_c-T}{T_c}\\
&\Delta C&=\frac{T_c\alpha_0^2}{2\beta}=n.
\end{eqnarray}

The effective Hamiltonian of noninteracting field of the order parameter fluctuations   is given by
\begin{equation}
H_0=\sum_{\bf k}\left (\alpha +\gamma_{ij}k_ik_j+2\pi k_z^2/k^2\right)
M_{\bf k}M_{-{\bf k}},
\end{equation}
where $M_{\bf k}=\int M({\bf r})e^{-i{\bf k}{\bf r}}d^3{\bf r}$.
The corresponding  free energy and the specific heat are\cite{StatPhys,Lev}
\begin{equation}
{\cal F}_{fl}=-\frac{T}{2}\sum_{\bf k}\ln\frac{\pi T}{\alpha +\gamma_{ij}k_ik_j+2\pi k_z^2/k^2},
\end{equation}
\begin{equation}
C_{fl0}=\frac{T^2\alpha_0^2}{2(2\pi)^3}\int\frac{dk_xdk_ydk_z}{[\alpha+2\pi \hat k_z^2+\gamma_{ij}k_ik_j]^2}. 
\end{equation}
Proceeding to spherical coordinates  and performing integration over modulus $k$ we come to
\begin{equation}
C_{fl0}=\frac{T^2\alpha_0^2}{32\pi^2}\int_0^1d\zeta\int_0^{2\pi}\frac{d\varphi}{(\alpha+2\pi \zeta^2)^{1/2}(\gamma_{\perp}+\zeta^2(\gamma_z-\gamma_{\perp}))^{3/2}}. 
\label{C}
\end{equation}
Here, $\gamma_{\perp}(\varphi)=\gamma_{x}\cos^2\varphi+\gamma_{y}\sin^2\varphi$.
At critical temperature $\alpha=0$ and the integral diverges. 
Hence, performing integration over $\zeta$ with logarithmic accuracy we obtain
\begin{equation}
C_{fl0}=\frac{T_c^2\alpha_0^2}{32\pi\sqrt{2\pi}\gamma^{3/2}}\ln\frac{\alpha}{2\pi}\approx\frac{n}{32\pi}\sqrt{\frac{T_c}{2\pi m^2n}}\ln\frac{2\pi m^2n}{T-T_c},
\label{CC}
\end{equation}
where $$\frac{1}{\gamma^{3/2}}=\frac{1}{2\pi}\int_0^{2\pi}\frac{d\varphi}{\gamma_\perp^{3/2}(\varphi)}.$$

The used condition $\alpha \ll 2\pi$ at $T_c=10K$ is realized at
\begin{equation}
\frac{T-T_c}{T_c}<\frac{2\pi m^2n}{T_c}\approx 0.015.
\label{T}
\end{equation}
In view of roughness of the parameter estimation the region of logarithmic increase of specific heat
can be in fact broader.

The  calculation taking into account the interaction of fluctuations has been performed by Larkin and Khmelnitskii\cite{Khm}.
In our notations
the expression  for the fluctuation specific heat at const pressure obtained in this paper is
\begin{equation}
C_{fl}=\frac{3^{1/3}T_c^2\alpha_0^2}{16\pi\gamma_{LK}^{2/3}\gamma^{3/2}}
\left (\ln\frac{\alpha}{2\pi}\right )^{1/3}
\label{Khm}
\end{equation}
Here $\gamma_{LK}=\frac{3T_c\beta}{\sqrt{32\pi}\gamma^{3/2}}$ is the effective constant of interaction.  Using the Eqs. (2)-(4) one can rewrite Eq. (\ref{Khm}) as
\begin{equation}
C_{fl}\approx\frac{n}{10}\left (\frac{T_c}{2\pi m^2n} \right)^{1/6}\left (\ln\frac{2\pi m^2n}{T-T_c}\right )^{1/3}.
\label{Khme}
\end{equation}
So, we see that the order parameter fluctuations give rise the increase of specific heat near the critical point. The power of the logarithm $(\ln\frac{\alpha}{2\pi})^{1/3}$ is quite slow function slightly exceeding unity, 
hence in the temperature region given by inequality (\ref{T}), and, perhaps,  even broader,
one may estimate the fluctuation specific heat as 
\begin{equation}
C_{fl}>\frac{n}{5}.
\label{Khmel}
\end{equation}

\section{Instability of the second order phase transition}

Larkin and Pikin \cite{Pikin} have pointed out that if the phase transition in a lattice with fixed volume is accompanied by an  increase in the specific heat  and the medium is isotropic in its elastic properties,  the second order type transition inevitably  changes to the first order one. There was found that a first-order transition will occur when
\begin{equation}
\frac{1}{T_c}\frac{4\mu K}{3K+4\mu}\left (\frac{\partial T_c}{\partial P}\right )^2C_P>1.
\label{LP}
\end{equation}
Here $K$ and $\mu$ are the nonsingular parts of the bulk and shear moduli and $C_P$ is the anomalous part of the specific heat.   
If the left hand side of Eq.(\ref{LP})  exceeds unity even if we take as $C_P$ the mean field jump in the specific heat $C_P=\Delta C$, then the first-order phase transition occurs \cite{Bean}. Otherwise we should take 
$C_P=C_{fl}$. In the latter case, 
usually, the left hand side in Eq.(\ref{LP}) is quite small
and the transition of the first order occurs at temperature $T^\star$ close to the critical temperature  where fluctuation specific heat is large enough. It means that the  temperature difference $T^\star-T_c$ is smaller than the critical temperature $T_c$ by many orders. The latent  heat at this transition 
\begin{equation}
q\approx C_{fl}(T^\star-T_c)
\end{equation}
proves to be extremely small. So, the first order phase transition is practically indistinguishable from the second order one and called weak first order phase transition or the phase transition of the first order closed to the second order.

Another source of instability towards to the first order transition in uniaxial ferroelectrics applicable as well to quite anisotropic ferromagnets was pointed out by Khmelnitskii and Shneerson \cite{Shneer} bearing in mind the anisotropy of elastic properties. In a media with anisotropic elastic properties the vertex of interaction between the critical fluctuations depends on the angles between the momentum-transfer vector and the crystal axes.  It leads to the development of instability: the amplitudes of interaction changes sign at finite value of $T-T_c$ and the first order type transition occurs. 
The latter type of instability can happen at higher temperature than the instability of the Larkin-Pikin due to  specific heat increase
 starts to be effective.  One may also imagine the opposite situation when the Larkin-Pikin type of instability occurs at higher temperature than the anisotropic fluctuation interaction begins to play significant role.
So, one can say that  in a concrete case: either the first order type transition occurs  at temperature  given by the Larkin-Pikin condition or it happens even at higher temperature by means of the Khmelnitskii-Shneerson mechanism.  Thus,  for quantitative estimation it is convenient to work with simple criterium  given by Eq. (\ref{LP}).

 In UGe$_2$  we deal with a particular situation. Namely, the Curie temperature in this compound falls monotonically with increasing pressure from 53 K at ambient pressure and drops precipitously above 15 Kbar. \cite{Saxena}  The average value of the critical temperature derivative can be estimated as 
\begin{equation}
 \frac{\partial T_c}{\partial P}\approx\frac{40~Kelvin}{14~kbar}=4\times 10^{-25}~cm^3
 \label{est}
\end{equation} 
 For the elastic moduli combination we have 
 \begin{equation}
 \frac{4\mu K}{3K+4\mu}\sim \rho c^2\approx 10^{11}erg/cm^3,
 \end{equation}
 where we have substituted  typical sound velocity  $c\approx10^5~cm/sec$ and used known \cite{Boulet} density value $\rho=10.26~g/cm^3$. Thus,  making use the estimation for the specific heat jump
 given by the Eqs. (6) (according to Eq.(\ref{Khmel}) the fluctuation specific heat play not dominant role here) we have for the combination Eq. (\ref{LP})
 
\begin{equation}
\frac{n}{T_c}\frac{4\mu K}{3K+4\mu}\left (\frac{\partial T_c}{\partial P}\right )^2=0.2.
\label{LPi}
\end{equation}
At $T\approx10K$  the pressure derivative of the critical temperature is much higher (and its square is even more higher) than its average value given by Eq. (\ref{est}). So, we come to conclusion 
that at critical temperature of the order  10 K the  Bean-Larkin-Pikin criterium is fulfilled.

\section{Conclusion}
The magneto-elastic interaction provides development of the first order instability at the phase transition to the ordered state in a ferromagnet. However,  actual temperature interval of this  instability development is negligibly small and the first order transition looks almost indistinguishable from the second order one.
The particular feature of anisotropic ferromagnet UGe$_2$ is the precipitous drop of the Curie temperature as the function of pressure near 14-15 kbar. Due to this property
at about these pressures the second order phase transition (or very weak transition of the first order)  to ferromagnet state turn into the real first order type transition.  

Under magnetic field parallel to the direction of spontaneous magnetization one should expect some increase of temperature of the first order instability in respect to its zero field value at the same pressure.

\acknowledgments

This work was partly supported by the grant SINUS of  Agence Nationale de la Recherche.

The author is indebted to V. Taufour for the  possibility to be familiar  with his  experimental results before its publication and the interest to the paper.

\appendix
\section{Comments to the paper\cite{Yamada}}

As it was mentioned in Introduction the treatment of the  $P-T-H$ phase diagram undertaken in the paper \cite{Yamada}  is characterized by several  ungainly features. Keeping the authors notations, here we point out two of them.

1. The fourth order term in the Landau expansion has the following form 
\begin{equation}
\frac{bm^4}{4}.
\end{equation}
The  temperature independent part of the coefficient $b=b(0)+ const ~T^2$
is strongly renormalized by magneto-elastic interaction
\begin{equation}
b(0)=b-2\kappa C_{mv}^2
\end{equation}
and has a negative value , that is
\begin{equation}
2\kappa C_{mv}^2>b.
\label{c}
\end{equation}
Here, $\kappa$ is the compressibility, and $C_{mv}$ is a magneto-volume coupling constant.
The author does  not prove the validity of this condition which is crucial  for the whole paper logic.
So, let us accept it as it is.

Then the author introduces parameter  $\eta$ determined by
\begin{equation}
\eta=\frac{1}{7}\frac{2\kappa C_{mv}^2}{|b-2\kappa C_{mv}^2|}
\label{e}
\end{equation}
and along whole article he  works with particular value $\eta=0.01$.  Substituting this value in the Eq.(\ref{e})
one can find relationship between $b$ and $2\kappa C_{mv}^2$ which obviously does not obey to the condition Eq. (\ref{c}). 

2. Following to the author,  let us  put 
\begin{equation}
\frac{b(0)}{|b(0)|}=-1
\end{equation}
and let us work with $\eta$ as with an independent parameter.
Then calculating the pressure derivative of the critical temperature  at tricritical point $P=P_t$ we find
 \begin{equation}
 \frac{\partial T_c}{\partial P}\propto -\frac{1}{2T_c(P_t)\eta(1-14\eta/5)}
 \end{equation}
Obviously this value is quite large at small $\eta$. According to Eq. (\ref{e}) in the absence of magneto-elastic interaction $\eta=0$. Hence, $\partial T_c/\partial P$ at tricritical point diverges
(tends to $-\infty$).
The presence of such a singular behavior
is physically senseless.

So, the model \cite{Yamada} is mathematically inconsistent  and leads to unphysical pressure dependence of critical temperature near the tricritical point.

\end{document}